\documentclass[twocolumn,prl,aps,groupedaddress,nopacs,footinbib,preprintnumbers,floats,floatfix]{revtex4}

\usepackage{amsmath}
\usepackage{amssymb}
\usepackage{latexsym}
\usepackage{graphicx}
\usepackage{hyperref}
\usepackage{bm}

\begin{document}

\title{Dynamical Dark Energy or Simply Cosmic Curvature?}

\author{Chris Clarkson$^1$, Marina Cort\^{e}s$^{1,2}$ and Bruce Bassett$^{1,3}$\\
\it $^1$Cosmology \& Gravity Group, Department Mathematics
and Applied Mathematics, University of Cape Town, Rondebosch 7701,
South Africa.\\ $^2$Astronomy Centre, University of Sussex, Brighton BN1 9QH, United
Kingdom\\$^3$ South African Astronomical Observatory, Observatory, Cape Town, South Africa}

\begin{abstract}
We show that the assumption of a flat universe induces critically large errors in
reconstructing the dark energy equation of state at $z \gtrsim 0.9$
even if the true cosmic curvature is very small, ${\cal O}(1\%)$ or
less. The spuriously reconstructed $w(z)$ shows a range of unusual behaviour, including crossing of the
phantom divide and mimicking of standard tracking quintessence models. For 1\% curvature and
$\Lambda$CDM, the error in $w$ grows rapidly above $z \sim 0.9$ reaching $(50\%,100\%)$ by redshifts of $(2.5,2.9)$ respectively, due to the long cosmological lever arm. Interestingly, the $w(z)$ reconstructed from distance data and Hubble rate measurements have opposite trends due to the asymmetric influence of the curved geodesics. These results show that including curvature as a free parameter is imperative in any future analyses attempting to pin down the dynamics of dark energy, especially at moderate or high redshifts.
\end{abstract}


\maketitle

\paragraph{\textbf{Introduction}}

The quest to distinguish between a cosmological constant, dynamical dark energy
and modified gravity has become the dominant obsession in cosmology. Formally elevated to the status of one of the most important problems in fundamental physics,
\footnote{http://www.ostp.gov/html/physicsoftheuniverse2.pdf} \footnote{http://www.pparc.ac.uk/roadmap/}
uncovering the true nature of dark energy, as encapsulated in the ratio of its pressure to
density, $w(z) = p_{DE}/\rho_{DE}$, has become the focus of multi-billion dollar proposed experiments using a
wide variety of methods, many at redshifts above unity (see e.g. \cite{DETF}).

Unfortunately these experiments will only measure a meagre number of $w(z)$
parameters to any precision -- perhaps two or three
\cite{numparam} \footnote{Note that non-standard methods such as varying-alpha offer the possibility of extracting significantly
more information \cite{VA}.} -- since the standard methods all involve integrals over $w(z)$ and typically suffer from subtle systematic effects.
As a result of this information limit, studies of dark energy have
traditionally fallen into two groups. The first group (see
e.g.~\cite{group1}) have taken their parameters to include $(\Omega_m,
\Omega_{DE}, w)$ with $w$ constant and often set to $-1$. The 2nd
group are interested in dynamical dark energy and have typically
assumed $\Omega_k = 0$ for simplicity while concentrating on constraining $w(z)$
parameters (see e.g. \cite{group2} ) \footnote {There is however a
small group of studies which have allowed both for dynamical
$w(z)$ and $\Omega_k \neq 0$; e.g. \cite{IT06, ZX06, wright}.}.

In retrospect, the origins of the common practise of ignoring
curvature in studies of dynamical dark energy are clear. Firstly,
the curved geodesics add an unwelcome complexity to the analysis
that has typically been ignored in favour of studies of different
parametrisations of $w(z)$. Secondly, standard analysis of the
Cosmic Microwave Background (CMB) and Baryon Acoustic Oscillations
(BAO) in the context of adiabatic $\Lambda$CDM also put stringent limits
on the curvature parameter, e.g. $\Omega_k = -0.003 \pm 0.010$
from WMAP + SDSS \cite{sdss06,dan}.
As a result it was taken for granted that the impact on the reconstructed
$w(z)$ would then be proportional to $\Omega_k$ and hence
small compared to experimental errors.

Further support for the view
that $\Omega_k$ should not be included in studies of dark energy
came from information criteria \cite{liddle} which showed that
for adiabatic {\em $\Lambda$}CDM including curvature is not
warranted from a model selection viewpoint.
Finally, even 2-parameter models of dark energy suffer from severe limitations \cite{cost}.
To include $\Omega_k$ as an additional parameter would only
further dilute constraints on $w(z)$.

However, we will show that ignoring $\Omega_k$
induces errors in the reconstructed dark energy equation of state, $w(z)$,
that grow very rapidly  with redshift and dominate the $w(z)$ error
budget at redshifts ($z \gtrsim 0.9$) even if $\Omega_k$ is very small. The aim of this paper is to argue that future studies of dark energy, and in particular, of observational data, should include $\Omega_k$ as a parameter to be fitted alongside the $w(z)$ parameters.

Looking back, this conclusion should not be unexpected. Firstly the case for flatness at the sub-percent level is not yet compelling: a general $\Lambda$CDM analysis \cite{dunkley}, allowing for general
correlated adiabatic and isocurvatrue perturbations, found that WMAP, together with large-scale structure and HST Hubble constant constraints, yields $\Omega_k = -0.06 \pm 0.02$. We will show that significantly smaller values of $\Omega_k$ lead to large effects at redshifts $z \gtrsim 0.9$ well within reach of the next generation of surveys.

Secondly, Wright (e.g. \cite{wright2}) has petitioned hard against the circular logic that
one can prove the joint statement $(\Omega_k = 0,w = -1)$ by simply proving the two conditional statements $(\Omega_k=0 | w = -1)$ and $(w = -1 | \Omega_k=0)$.
This has been verified in \cite{IT06, wright} where values of $|\Omega_k| = 0.05$ or larger are found to be acceptable at $1-\sigma$ if one allows for 2-parameter varying $w(z)$. Clearly with more dark energy parameters - or correlated isocurvature perturbations - even larger $\Omega_k$ would be consistent with the current datasets. Given that the constraints on $\Omega_k$ evaporate precisely when $w$ deviates most strongly from a cosmological constant, it is clearly inconsistent to assume
$\Omega_k = 0$ when deriving constraints on dynamical dark energy. The uncertainty around the current value of $\Omega_k$ begs the question, how does the error on $w(z)$ scale with $\Omega_k$?

We show below and in Figure~1 that the growth of
the error in $w(z)$, arising from erroneously assuming the cosmos to be flat,
is very rapid, both in redshift and $\Omega_k$.
The reason for this rapid  growth of the
error in $w$ can be easily understood. Consider a $\Lambda$CDM model
with small but non-zero $\Omega_k$. The curvature contribution to
$H(z)^2$ scales as $\Omega_k (1+z)^2$ while $\Omega_{\Lambda}$
is constant. In addition the curved geodesics drastically alter observed
distances when they are a sizeable fraction of the curvature radius.
If one tries to reproduce the observations in such a universe with
a flat model with varying $w(z)$, it is clear that $w(z)$ must
deviate strongly from $w = -1$ as $z$ increases as the dark energy
tries to mimic the effects of the rapidly growing curvature {\em
and} the curved geodesics. A crude estimate for the redshift for strong deviations
from the true $w$ would follow from setting $\Omega_k (1+z)^2 \sim \Omega_{DE}$.
For values of $\Omega_k = 0.05$ and $\Omega_{DE}  = 0.7$ this
yields $z \approx 2.7$, within reach of BAO surveys such as
WFMOS \cite{wfmos} and VIRUS\footnote{http://www.as.utexas.edu/hetdex/}.

We will show in detail below that the actual redshift where
problems set in is significantly lower in the case of luminosity
distance measurements due to the added curved geodesic effects.
In fact, if the curvature is negative, a redshift is reached where
the dark energy cannot mimick the curvature at all, unless $\rho_{DE}$ can
change sign, and the reconstructed dark energy has $w \rightarrow -\infty$. Related
work \cite{linder} has investigated some of these issues. However because a restricted parameterisation
was assumed for $w(z)$ the results found were significantly less severe than we find.

\paragraph{\textbf{Reconstructing the DE equation of state from observations}}

Here we wish to compare the reconstructed dark energy $w(z)$ from perfect distance (either luminosity, $d_L(z)$, or area, $d_A(z)$) and Hubble rate ($H(z)$) measurements as these are the basis for all modern cosmology experiments including those using Type Ia supernovae, BAO, weak lensing etc...
Since we are interested in the effects of ignoring curvature, let us assume we have perfect knowledge of both $H(z)$ and $d_L(z)$ across a
range of redshifts: how would we reconstruct $w(z)$? The luminosity distance may be written as
\begin{equation}\label{d_L}
d_{L}(z)=\frac{c(1+z)}{H_0 \sqrt{-\Omega_k}}\sin{\left(
\sqrt{-\Omega_k}\int_0^z{\mathrm{d}z'\frac{H_0}{H(z')}}\right)}
\end{equation}
which is formally valid for all curvatures, where $H(z)$ is given by the Friedmann equation,
\begin{widetext}
\vspace*{-5mm}
\begin{eqnarray}
H(z)^2&=& H_0^2\biggl\{\Omega_{m} (1+z)^3+\Omega_{
k}(1+z)^2\nonumber
+\Omega_{DE}\exp{\left[3\int_0^z
\frac{1+w(z')}{1+z'}\mathrm{d}z'\right]}\biggr\},
\label{hubble}
\end{eqnarray}
and $\Omega_{DE}=1-\Omega_m-\Omega_k$.
Knowledge of $H(z)$ allows us to directly probe the dynamics of the universe, while $d_L(z)$ is strongly affected by the cosmic curvature which distorts null geodesics away from straight lines.

We can invert these functions to give two different expressions for the equation of state of the dark energy, $w(z)$: one in terms of $H(z)$ and its first derivative,
\begin{equation}\label{wH}
w(z)=-\frac13\frac{\Omega_k H_0^2(1+z)^2+2(1+z)HH'-3H^2}{
H_0^2(1+z)^2[\Omega_m(1+z)+\Omega_k]-H^2} \, ,
\end{equation}
where $'=\mathrm{d}/\mathrm{d}z$; and the other in terms of the the dimensionless luminosity distance ${D}_L=(H_0/c) d_L$, and its derivatives
\begin{equation}\label{wdL}
w(z)=\frac{2}{3}\frac{(1+z)\left\{[\Omega_{ k} {D}_{ L}^2+(1+z)^2]{D}_{
L}''-\frac{1}{2}(\Omega_{ k} {D}_{
L}'^2+1)[(1+z){D}_{ L}'-{D}_{ L}]\right\}}
{[(1+z){D}_{
L}'-{D}_{ L}]\left\{(1+z)[\Omega_{ m}(1+z)+\Omega_k]{D}_{
L}'^2-2[\Omega_{ m} (1+z)+\Omega_{ k}] {D}_{ L}{D}_{
L}'+\Omega_{ m} {D}_{ L}^2-(1+z)\right\}}\,.
\end{equation}
Note that $\Omega_{DE}$ has dropped out of both expressions. \\[0mm]

\end{widetext}
\vspace*{-10mm}


\begin{figure*}[p!]
\includegraphics[width=0.95\textwidth]{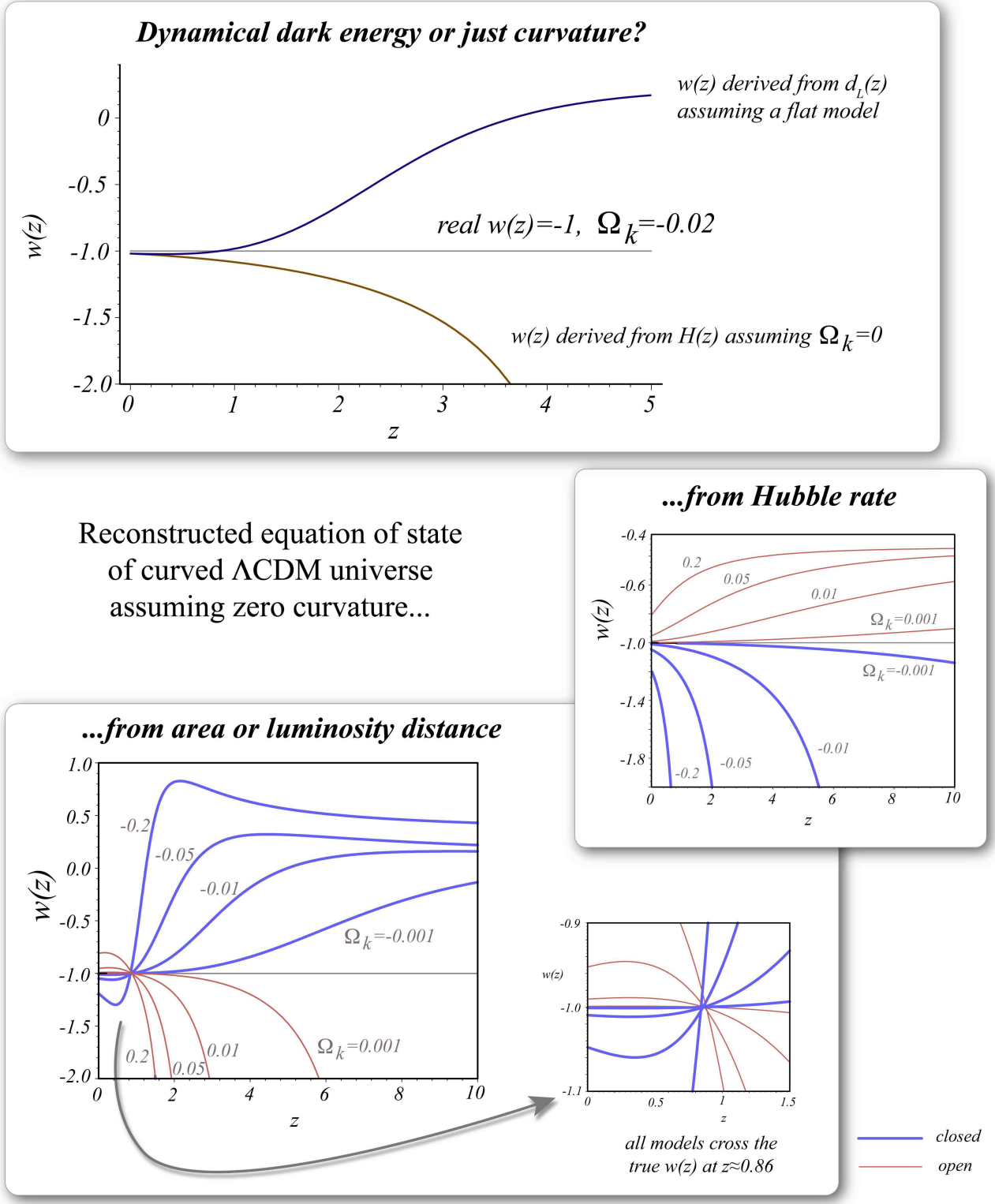}
\caption{{\bf Reconstructing the dark energy equation of state assuming zero curvature} when the true curvature is 2\% in a closed $\Lambda$CDM universe. The $w(z)$ reconstructed from $H(z)$ is phantom ($w < -1$) and rapidly acquires an error of order 50\% and more at redshift $z\gtrsim2$, and diverges at finite redshift.  The reconstructed $w(z)$ from $d_L(z)$ for $\Omega_k<0$ is phantom until $z\approx0.86$, where it crosses the true value of $-1$ and then crosses 0 at high redshift, where the bending of geodesics takes over from dynamical behavior, producing errors in opposite direction to the DE reconstructed from $H(z)$.
In order to make up for the missing curvature, the reconstructed DE is
behaving like a scalar field with a tracking behavior. These effects arise even if the curvature is extremely small $<0.1\%$.}
\end{figure*}


If, in addition, we also knew $\Omega_m$ and $\Omega_k$ perfectly then these two expressions would yield the same function $w(z)$. But what if -- as is commonly assumed -- we impose $\Omega_k = 0$ when in fact the true curvature is actually non-zero? It is usually implicitly assumed that
the error on $w(z)$ will be of order $\Omega_k$, and this is indeed true for $z \lesssim 0.9$ (see Figure 1). However, this intuition is strongly violated  for $z \gtrsim 0.9$, even given perfect knowledge of $d_L(z)$ or $H(z)$.

\paragraph{\textbf{Zero curvature assumption}}

We may uncover the implications of incorrectly assuming flatness by computing the $w=-1, \Omega_k\neq 0$ functional forms for $d_L(z)$ and $H(z)$ and inserting the results into Eqs (\ref{wH}) and (\ref{wdL}). If we then set $\Omega_k=0$ in Eqs. (\ref{wH}) and (\ref{wdL}) we arrive at the two corresponding $w(z)$ functions (if they exist) required to reproduce the curved forms for $H(z)$ and $d_L(z)$ [or $d_A(z)$ -- for any distance indicator the results are exactly the same]. Figure 1 presents for simplicity the concordance value of $w=-1$ but we have checked that the qualitative results do not depend on the `true' underlying dark energy model \footnote{In fact, the results presented here are qualitatively the same for any assumed $\Omega_k$ which is different from the true value.}. We assume $\Omega_m=0.3$ in all expressions\footnote{There is a degenracy between $\Omega_m$ and dark energy in general \cite{m1}}; numbers quoted are weakly dependent on this. The resulting $w(z)$ can then be thought of as the function which produces the same $H(z)$ or $d_L(z)$ as the actual curved $\Lambda$CDM model: e.g.,
\begin{equation}
d_L[\text{flat},w(z)]=d_L[\text{curved},w(z)=-1].
\end{equation}

In the figure we show what happens for $\Lambda$CDM: curvature manifests itself as evolving dark energy. In the case of the Hubble rate measurements this is fairly obvious: we are essentially solving the equation $\Omega_{DE} f(z) = \Omega_{\Lambda} + \Omega_k (1+z)^2$ where $f(z)$ is the integral in the last term of Eq. (\ref{hubble}). For $\Omega_k > 0$, $w(z)$ must converge to $-1/3$ to compensate for the curvature. For $\Omega_k < 0$, the opposite occurs and a redshift is reached when $w \rightarrow -\infty$ in an attempt to compensate (unsuccessfully) for the negative curvature. Already we can see why the assumption that the error in $w$ is of order the error in $\Omega_k$ breaks down so drastically.

Interestingly, the curved geodesics imply that the error in $w$ reconstructed from $d_L(z)$ and $H(z)$ have opposing signs at $z \gtrsim 0.9$, as can be seen by comparing the panels in Fig. 1. Above the critical redshift the effect of curvature on the geodesics becomes more important than the pure dynamics, and the luminosity distance flips $w(z)$ in the opposite direction to that reconstructed from $H(z)$.
Given that dark energy reconstructed from $H(z)$ pulls in the other direction implies that knowledge of $H(z)$ is a powerful tool in measuring curvature. In fact, we may directly combine measurements of $H(z)$ and $D_L(z)$ to measure curvature independently of other cosmological parameters (except $H_0$) via
\begin{equation}
\Omega_k=\frac{\left[H(z)D'(z)\right]^2-c^2}{[H_0D(z)]^2}.
\end{equation}
where $d_L(z)=(1+z)D(z)$.
This is an advantage of BAO surveys which provide a simultaneous measurement of both $d_A$ and $H(z)$ at the same redshift and, with Lyman-break galaxies as targets, can easily reach $z = 3$ \cite{wfmos}.

\paragraph{\textbf{Conclusions}}

\begin{figure}[htbp]
\begin{center}
\includegraphics[width=\columnwidth]{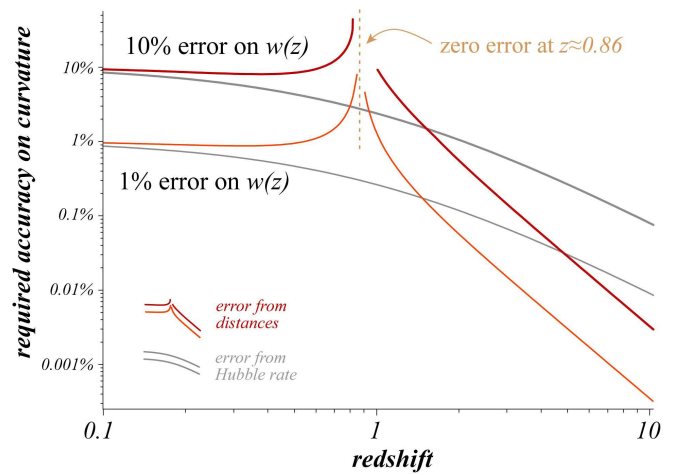}
\caption{The accuracy with which we must know $\Omega_k$ in order to measure $w(z)$ to within a given error.}
\label{fig2}
\end{center}
\end{figure}

We have argued that cosmic curvature must be included as a free
parameter in all future studies of dark energy. This is particularly
crucial at redshifts $z \gtrsim 0.9$ where the resulting error in $w(z)$ can
exceed $\Omega_k$ by two orders of magnitude or more. Even with perfect measurements
of $d_L(z)$ or $H(z)$ the error induced on the
measured $w(z)$ by assuming flatness when in fact $\Omega_k =
0.01$ reaches $100\%$ by $z \approx 2.9$; sub-percent errors on $w(z\sim5)$
would require $|{\Omega_k}|\lesssim10^{-5}$~-- see Figure~\ref{fig2}. This implies that we may \emph{never} do better than 1\% errors in $w(z)$ since we expect $\Omega_k\sim10^{-5}$, even if inflation predicts zero curvature, at the level of perturbations.
It is interesting to note that the
backreaction of cosmological fluctuations may cause effective non-zero curvature
that may yield practical limits on our ability to measure $w(z)$ accurately at
$z > 1$ (see e.g. \cite{backreact}).

Finally it is interesting to ask what the results imply for other tests of dark energy such as the
integrated Sachs-Wolfe (ISW) effect. Since the ISW effect is sensitive to both curvature and dark energy, and since
it is possible to probe the ISW effect at $z > 1$ \cite{isw} the effect of curvature may also be important there. Similar
sentiments apply to cluster surveys which are sensitive to both the growth function and the volume of space and can detect clusters
at high redshift through the Sunyaev-Zel'dovich effect. Since  the volume of space is
very sensitive to $\Omega_k$ (see e.g. \cite{open}) it is likely that the reconstructed $w(z)$ from this
method will also be subject to large errors if one incorrectly assumes flatness.

As a result, measuring $\Omega_k$ accurately (to $\sigma_{\Omega_k} < 10^{-3}$) will play an
important role in the quest to hunt down the true origin of dark energy in the coming decade.
Baryon Acoustic Oscillations, in conjunction with other probes of distance
such as weak lensing are likely to play a key role in illuminating these shadows of curvature \cite{fc}.

{\em Acknowledgements -- }we thank Chris Blake, Thomas Buchert, Daniel Eisenstein, George Ellis, Ren\'ee Hlozek,
Martin Kunz, Roy Maartens, Bob Nichol and David Parkinson for useful comments and insights. MC thanks
Andrew Liddle and FTC for support. BB and CC acknowledge support from the NRF.



\begin{references}

\bibitem{DETF}A. Albrecht {\em et al.}, Report of the Dark Energy Task Force, astro-ph/0609591 (2006)

\bibitem{VA} See e.g. T. Chiba, K. Kohri, Prog.Theor.Phys. {\bf 107} (2002) 631; D.~Parkinson, B.~A.~Bassett and J.~D.~Barrow,
  Phys.\ Lett.\  B {\bf 578} (2004) 235; C. Wetterich, JCAP {\bf 0310}, 002 (2003);  T. Damour, F. Piazza, G. Veneziano
Phys.Rev.Lett. {\bf 89},  081601 (2002)

\bibitem{numparam} E. Linder and D. Huterer, Phys.Rev. D{\bf 72} 043509, (2005)

\bibitem{IT06} K. Ichikawa, T. Takahashi, (2006) astro-ph/0612739

\bibitem{ZX06}G-B. Zhao, J-Q. Xia, H. Li, C. Tao, J-M.Virey, Z-H. Zhu, X. Zhang, (2006), astro-ph/0612728

\bibitem{wright} E.~L.~Wright,
  arXiv:astro-ph/0701584.


\bibitem{linder} E. V. Linder, Astropart.Phys. {\bf 24},  391 (2005); D. Polarski, A. Ranquet, Phys Rev Lett. B {\bf627}, 1 (2005)

\bibitem{liddle}
A.~R.~Liddle,
Mon.\ Not.\ Roy.\ Astron.\ Soc.\  {\bf 351}, L49 (2004)

\bibitem{dunkley}
J. Dunkley {\em et al.}, Phys.Rev.Lett. {\bf 95} (2005) 261303

\bibitem{group1}
%
See e.g. P.~Astier {\it et al.},
  Astron.\ Astrophys.\  {\bf 447}, 31 (2006);
  C.Zunckel, P.G Ferreira, astro-ph/0610597; D. N. Spergel {\em et al}., astro-ph/0603449

\bibitem{group2}
See e.g.
  B.~A.~Bassett, M.~Kunz, J.~Silk and C.~Ungarelli,
  Mon.\ Not.\ Roy.\ Astron.\ Soc.\  {\bf 336}, 1217 (2002);
  P.~S.~Corasaniti {\em et al.},
  Phys.\ Rev.\ Lett.\  {\bf 90}, 091303 (2003);
  P.~S.~Corasaniti {\em et al.},
  Phys.\ Rev.\  D {\bf 70}, 083006 (2004);
  D. Rapetti, S. W. Allen, J. Weller, MNRAS, {\bf 360}, (2005) 555;
  S. Hannestad, E. Mortsell, JCAP {\bf 0409} (2004) 001;
  M.~Doran, G.~Robbers and C.~Wetterich,
  Phys.\ Rev.\  D {\bf 75}, 023003 (2007);
  H. K. Jassal, J. S. Bagla, T. Padmanabhan, astro-ph/0601389;
  W.~M.~Wood-Vasey {\it et al.},
  arXiv:astro-ph/0701041.


\bibitem{sdss06}
 M.~Tegmark {\it et al.},
  Phys.\ Rev.\  D {\bf 74}, 123507 (2006)

\bibitem{dan}
D. J. Eisenstein {\em et al.}, Ap.J. {\bf 633} (2005) 560

\bibitem{cost}
B.~A.~Bassett, P.~S.~Corasaniti and M.~Kunz,
  Astrophys.\ J.\  {\bf 617}, L1 (2004)


\bibitem{wright2}
E.~L.~Wright,
  arXiv:astro-ph/0603750, (2007)

\bibitem{m1}
M. Kunz, astro-ph/0702615, (2007)

\bibitem{wfmos}
K. Glazebrook and the WFMOS Feasibility Study Dark Energy Team,
White paper submitted to the Dark Energy Task Force, astro-ph/0507457;
B.~A.~Bassett, R.~C.~Nichol and D.~J.~Eisenstein  [for the WFMOS Collaboration],
arXiv:astro-ph/0510272;

\bibitem{backreact}
See e.g. T. Buchert, M. Carfora, Phys. Rev. Lett. {\bf 90}, 031101 (2003);
N. Li, D. J. Schwarz, gr-qc/0702043; S. Rasanen, Class. Quant. Grav. {\bf 23} (2006) 1823; A.A. Coley, N. Pelavas
Phys.Rev. D{\bf 75},  043506 (2007).

\bibitem{isw} T. Giannantonio {\em et al.}, Phys.Rev. D{\bf 74}, 063520 (2006)

\bibitem{open} M. Kamionkowski, D. N. Spergel, N. Sugiyama, Ap.J. {\bf 426} (1994) L57;
M. Kamionkowski, D. N. Spergel, Ap.J. {\bf 432}, 7 (1994)

\bibitem{fc} G. Bernstein, Astrophys.J. {\bf 637} (2006) 598; P. McDonald
and D.J. Eisenstein, astro-ph/0607122 (2006); L. Knox, Y-S. Song, H. Zhan, astro-ph/0605536 (2006); K. Glazebrook, C. Blake, Ap.J. {\bf 631}, 1 (2005)


\end{references}
\end{document}